\documentclass[twocolumn,letter]{jpsj2}

\usepackage{amsmath}
\usepackage{amsfonts}
\usepackage{amssymb}
\usepackage{txfonts}
\usepackage{bm}
\usepackage{tabularx}
\usepackage{graphicx,color}

\def\Vec#1{\bm{#1}}
\def\Hc2{H_\mathrm{c2}}
\def\Tc{T_\mathrm{c}}

\def\214{\mathrm{Sr_2RuO_4}}
\def\327{\mathrm{Sr_3Ru_2O_7}}
\title{Large Enhancement of 3-K Phase Superconductivity \\ in the Sr$_2$RuO$_4$-Ru Eutectic System by Uniaxial Pressure}

\author{
	Shunichiro \textsc{Kittaka}$^{1,\thanks{E-mail: kittaka@scphys.kyoto-u.ac.jp}}$, 
	Hiroshi \textsc{Yaguchi}$^{1,2}$,
	and Yoshiteru \textsc{Maeno}$^{1}$ 
}

\inst{$^{1}$Department of Physics, Graduate School of Science,
		Kyoto University, Kyoto 606-8502\\
	$^{2}$Department of Physics, Faculty of Science and Technology, 
		Tokyo University of Science, Noda, Chiba 278-8510
}

\recdate{\today}

\abst{
While the superconducting transition temperature $\Tc$ of $\214$ is 1.5~K, 
its onset $\Tc$ is enhanced as high as 3~K in the $\214$-Ru eutectic system,
which is often referred to as the 3-K phase.
We have investigated effects of uniaxial pressure on the non-bulk superconductivity in the 3-K phase.
While $\Tc$ of pure $\214$ is known to be suppressed by hydrostatic pressure,
a large enhancement of the superconducting volume fraction of the 3-K phase was observed for both out-of-plane and in-plane uniaxial pressures. 
Especially, under the in-plane pressure, the shielding fraction at 1.8~K of only less than 0.5\% at 0 GPa exceeds 40\% at 0.4~GPa.
Such a large shielding fraction suggests that 
under the uniaxial pressure interfacial 3-K phase superconductivity penetrates deep into the bulk of $\214$.
The present finding provides a significant implication to the unresolved origin of the enhancement of $\Tc$ to 3~K in the $\214$-Ru eutectic system.
}

\kword{Sr$_2$RuO$_4$, uniaxial pressure, spin-triplet superconductivity, ruthenate, eutectic system}

\begin{document}
\maketitle

The $n=1$ member of the Ruddlesden-Popper (R-P) type ruthenates Sr$_{n+1}$Ru$_n$O$_{3n+1}$, $\214$, 
is well established to be a spin-triplet superconductor.\cite{Maeno1994Nature,Mackenzie2003RMP}
Among a number of remarkable features in $\214$, 
an enhancement of the superconducting transition temperature $\Tc$ in the $\214$-Ru eutectic system,
which is often referred to as the ``3-K phase", is rather striking. 
Although the original superconducting phase in pure Sr$_2$RuO$_4$ occurs with a sharp transition at $\Tc$ of 1.5~K,
the $\214$-Ru eutectic system exhibits a broad superconducting transition with an enhanced onset $\Tc$ of approximately 3~K.\cite{Maeno1998PRL}
Several experimental facts suggest that the superconductivity with the enhanced $\Tc$ 
occurs on the $\214$ side of the $\214$-Ru interface and consists of filamentary loops among different Ru inclusions. \cite{Maeno1998PRL,Ando1999JPSJ,Yaguchi2003PRB,Kittaka2009JPSJ}
In fact, zero bias conductance peaks, a hallmark of unconventional superconductivity,
have been observed in tunneling measurements on S/N junctions at the interfaces. \cite{Mao2001PRL,Kawamura2005JPSJ,Yaguchi2006JPSJ}

While the origin of the enhancement of $\Tc$ in the 3-K phase remains uncertain, 
important aspects of its superconductivity were successfully described using a phenomenological theory 
within the framework of Ginzburg Landau formalism which assumes spin-triplet pairing similar to $\214$. \cite{Sigrist2001JPSJ,Matsumoto2003JPSJ}
Although the basic form of the vector order parameter of $\214$ in zero field is believed to be 
$\Vec{d}(k)=\Hat{z}\Delta_0(k_x \pm ik_y), $\cite{Mackenzie2003RMP}
it has been proposed that, in the 3-K phase, the degeneracy of the two components of the superconducting order parameter in $\214$
is lifted by broken tetragonal symmetry in $\214$ at the interface between $\214$ and Ru.
Only the component parallel to the interface would be stabilized at 3 K and 
the other component with a relative phase of $\pi$/2 emerges at a lower temperature.

It is known that the electronic states of the R-P type ruthenates are significantly affected by 
the rotation, tilting and flattening of RuO$_6$ octahedra,\cite{Matzdorf2000Science,Friedt2001PRB,Fang2001PRB} 
making uniaxial pressure an effective tool to control their electronic states.
For example, the metamagnetic normal metal $\327$, the $n=2$ member of the R-P series, 
exhibits ferromagnetism under uniaxial pressure along the [001] axis. \cite{Ikeda2004JPSJ,Yaguchi2006AIP}
In the present work, we have investigated uniaxial-pressure effects on the 3-K phase 
to obtain insight into the mechanism of the enhancement of its $\Tc$.
We have measured the dc magnetization of the $\214$-Ru eutectic system under uniaxial pressure along the [001], [100] and [110] axes, and
revealed that the superconducting volume fraction of the 3-K phase is strongly enhanced for pressure along all axes
while its onset $\Tc$ remains nearly the same.
Especially, under in-plane uniaxial pressure the shielding fraction at 1.8~K exceeds 40\%. 
This large shielding fraction suggests that 
interfacial 3-K phase superconductivity penetrates deep into the bulk of $\214$ under the uniaxial pressure.

\begin{figure}
\begin{center}
\includegraphics[width=2.3in]{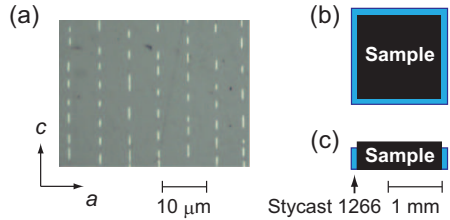}
\end{center}
\caption{
(Color Online) 
(a) Optical microscope photograph of the (100) surface of a $\214$-Ru eutectic crystal used in the present study. 
The dark (bright) parts correspond to Sr$_{2}$RuO$_{4}$ (Ru inclusions).
The three dimensional configuration of the Ru lamellae is expected to be similar to that illustrated in Fig.~1 of Ref.~6.
Schematic diagrams of (b) top and (c) side views of the sample surrounded by epoxy (Stycast 1266).
}
\label{sample}
\end{figure}

Measurements were performed on more than ten eutectic samples from four different batches grown by a floating zone method. \cite{Mao2000MRB}
Approximate dimensions of the samples were 1.5 $\times$ 1.5 $\times$ 0.3 mm$^3$.
These samples were cut and polished such that the shortest dimension was parallel to the [001], [100] or [110] axis 
(determined from Laue pictures).
As exemplified in Fig.~\ref{sample}(a),
we identified the orientation of Ru lamellae on the top and bottom surfaces on which the uniaxial pressure was applied.
The three-dimensional configuration of the lamellae is expected to be similar to that illustrated in Fig.~1 of Ref.~6 
with photographs of three orthogonal surfaces.
Typical dimensions of Ru inclusions are $10 \times 10 \times 1$ $\muup$m$^3$. 
Uniaxial pressure was applied parallel to the shortest dimension of each sample using a piston-cylinder type pressure cell
made of BeCu with a cylindrical outer body made of oxygen-free copper. 
In this cell, the pressure is maintained by dish-shaped springs made of BeCu.
Applied pressures were calculated from the forces applied to the samples at room temperature, 
which were confirmed to show a reasonable agreement with low-temperature pressure 
determined from the $\Tc$'s of tin and lead \cite{Smith1967PR}. 
In our previous study, \cite{Yaguchi2009JPCS}
we were not able to obtain quantitatively reproducible data 
because the applied force was partially released by a breakdown of the sample at high uniaxial pressures.
In order to maintain the applied pressure, 
side surfaces of the sample were covered with thin epoxy (Stycast 1266, Emerson \& Cuming), as depicted in Figs.~\ref{sample}(b) and \ref{sample}(c),
after the top and bottom surfaces were polished.
This made the data both qualitatively and quantitatively reproducible. 
In this paper, we focus on three eutectic samples
for identifying uniaxial-pressure effects along the [001], [100] and [110] axes of $\214$ from magnetization measurements. 
A SQUID (superconducting quantum interference device) magnetometer (MPMS, Quantum Design) was used to 
measure the total dc magnetization $M$ of the sample and pressure cell down to 1.8~K. 
The SQUID measurements were performed in the order of increasing applied pressure for each sample
with a dc magnetic field applied parallel to the direction of the applied pressure. 
Although the dc 
Meissner 
fraction for field cooling was about half of that for zero field cooling (ZFC), 
their dependences on uniaxial pressure and temperature were qualitatively the same.
For this reason, we have taken more extensive data for ZFC which yielded stronger signals.
In this paper, we present these ZFC data.
In addition, we present the specific heat of one of the eutectic samples without epoxy reinforcement. 
The specific heat $C_p$ was measured by a thermal relaxation method with a commercial calorimeter (PPMS, Quantum Design) down to 0.35~K 
on a sample after the pressure was released.

\begin{figure}
\begin{center}
\includegraphics[width=1.9in]{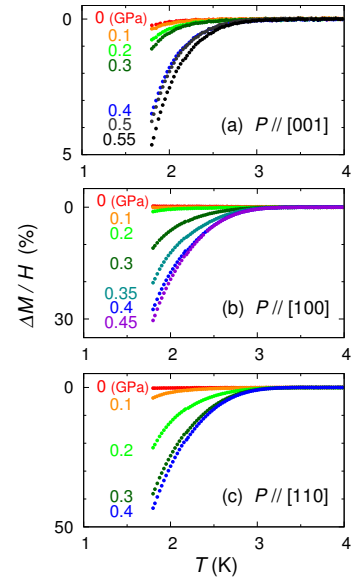}
\caption{
(Color Online) Temperature dependence of the dc shielding fraction of Sr$_{2}$RuO$_{4}$-Ru 
in a field of 2 mT (ZFC) at different uniaxial pressures: 
(a) parallel to the $c$ ([001]) axis, 
(b) parallel to the $a$ ([100]) axis, and
(c) parallel to the [110] direction.
Uniaxial pressures in GPa are shown.
}
\label{up}
\end{center}
\end{figure}

Figures~\ref{up}(a)-\ref{up}(c) show the temperature dependence of the dc shielding fraction 
$\Delta M/H$ at 2~mT under uniaxial pressure parallel to the [001], [100] and [110] axes, respectively.
In these figures, $\Delta M/H$ is normalized by the ideal value 
calculated for the full Meissner state without a demagnetization correction.
Note that the demagnetization effect leads to an apparent enhancement 
in the dc shielding fraction in the present case:
$\Delta M/H$ of lead with $\Tc=7.2$~K, whose dimensions are $2 \times 2 \times 0.2$ mm$^3$,
reached approximately 580\% (110\%) in the full Meissner state 
under a magnetic field applied parallel (perpendicular) to the shortest dimension,
which is estimated to be 640\% (110\%) using the calculated demagnetization factor.
However, it is not certain how large the demagnetization factor should effectively be 
since superconductivity in the 3-K phase is not uniform.

\begin{figure}
\begin{center}
\includegraphics[width=2in]{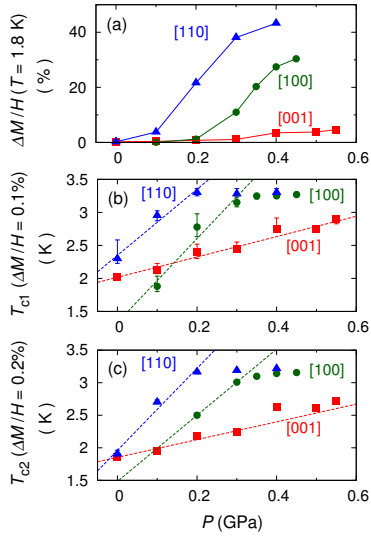}
\caption{
(Color online) 
(a) Uniaxial-pressure dependence of the dc shielding fraction $\Delta M/H$ at 1.8~K. The solid lines are guides to the eye.
Figures (b) and (c) represent uniaxial-pressure dependence of $T_\mathrm{c1}$ and $T_\mathrm{c2}$, 
at which $\Delta M/H$ becomes 0.1\% and 0.2\%, respectively.
The squares, circles and triangles represent the results for $P_{\parallel [001]}$, $P_{\parallel [100]}$ and $P_{\parallel [110]}$, respectively.
The dashed lines are $P$-linear fits to $\Tc(P)$ at low pressure.
}
\label{com}
\end{center}
\end{figure}

The dc shielding fractions at 1.8 K and 0~GPa are less than 0.5\% for all three samples. 
As shown in Figs.~\ref{up}(a)-\ref{up}(c), 
the application of uniaxial pressure in all three directions enhances the shielding fraction; 
however, the strength of the effect differs significantly.
We plot in Fig.~\ref{com}(a) the uniaxial-pressure dependence of the dc shielding fraction at 1.8~K 
for pressure along the [001], [100] and [110] axes. 
At a pressure of 0.4 GPa, these values increase to above 30\% for $P_{\parallel [100]}$ and $P_{\parallel [110]}$, and 
5\% for $P_{\parallel [001]}$
(note that the vertical scale in Fig. \ref{up}(c) is ten times larger than that in Fig.~\ref{up}(a)).
These results were reproducible for the other samples; 
the maximum shielding fraction never exceeded 10\% in an out-of-plane pressure of 0.4~GPa, 
while it greatly exceeded 10\% in an in-plane pressure of 0.4~GPa.

In order to characterize the uniaxial-pressure dependence of $\Tc$,
we here define the temperatures $T_\mathrm{c1}$ and $T_\mathrm{c2}$ at which $\Delta M/H$ becomes 0.1\% and 0.2\%, respectively.
The uniaxial-pressure dependences of $T_\mathrm{c1}$ and $T_\mathrm{c2}$ are plotted in Figs.~\ref{com}(b) and \ref{com}(c).
The uniaxial-pressure coefficients of $\Tc$, 
$\mathrm{d}\Tc/\mathrm{d}P_{\parallel [001]}$, $\mathrm{d}\Tc/\mathrm{d}P_{\parallel [100]}$ and $\mathrm{d}\Tc/\mathrm{d}P_{\parallel [110]}$, 
are approximately 1.5, 6.3 and 5.0 K/GPa for $T_\mathrm{c1}(P)$,
and 1.4, 5.1 and 6.3 K/GPa for $T_\mathrm{c2}(P)$, respectively, based on the initial linear slopes.
Despite unavoidable variations of $T_\mathrm{c1}$ and $T_\mathrm{c2}$ among different crystals,
we can conclude that the enhancement under in-plane pressure is greater by about a factor of four than that under out-of-plane pressure.

\begin{figure}
\begin{center}
\includegraphics[width=3in]{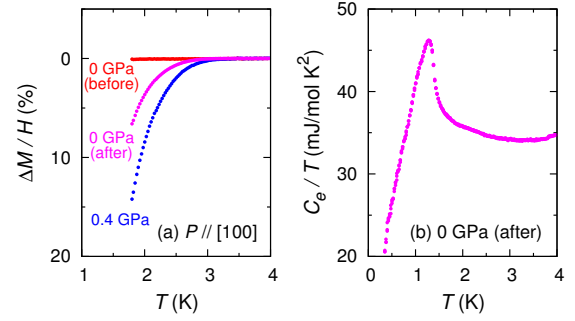}
\caption{
(Color online) 
(a) Temperature dependence of the dc shielding fraction of a sample without epoxy reinforcement measured in a field of 2 mT for $P \parallel [100]$.
(b) Temperature dependence of the electronic specific heat divided by temperature $C_\mathrm{e}/T$ of the same sample 
after releasing the uniaxial pressure of 0.4~GPa for $P_{\parallel [100]}$.
}
\label{Cp}
\end{center}
\end{figure}

In order to evaluate the enhancement of the superconducting volume fraction in the 3-K phase,
the specific heat of a eutectic sample without epoxy reinforcement was measured 
after releasing the applied uniaxial pressure $P_{\parallel [100]}$ of 0.4~GPa.
As shown in Fig.~\ref{Cp}(a), at 0.4~GPa this sample exhibits a dc shielding fraction of 14\% at 1.8~K,
compared with 28\% for the epoxy reinforced sample in Fig.~\ref{up}(b).
After the pressure was released, the sample retained a shielding fraction of as much as 7\%.
The electronic specific heat divided by temperature $C_\mathrm{e}/T$ is plotted in Fig.~\ref{Cp}(b). 
A sharp peak is observed at 1.3~K, which is attributed to the bulk superconductivity of $\214$.
In addition, we found 
an additional contribution above 1.5~K, exceeding 10\% of the electronic specific heat of the normal state.
This contribution is attributable to the 3-K phase superconductivity. 
It is at most 1\% at 0~GPa before applying pressure.\cite{Yaguchi2003PRB}
These results suggest that the superconducting volume fraction in the 3-K phase is strongly enhanced under uniaxial pressure.
In fact, the discontinuity at the bulk $\Tc$ of 1.3~K is suppressed to be nearly half of that at 0~GPa before applying pressure
($\Delta C_\mathrm{e}/T \sim 19$ mJ/mol K$^2$),\cite{Yaguchi2003PRB} 
which suggests entropy release also associated with superconductivity above the $\Tc$ of $\214$. 

While the results of uniaxial-pressure experiments on $\214$ have not been reported, 
the hydrostatic-pressure coefficient of $\Tc$, ${\rm d}T_{\rm c}/{\rm d}P$, for $\214$ 
has been estimated to be approximately $-0.2$~K/GPa.\cite{Shirakawa1997PRB,Forsythe2002PRL}
Because ${\rm d}T_{\rm c}/{\rm d}P$ is negative, 
the basic relation in tetragonal symmetry, 
${\rm d}\Tc / {\rm d}P =2{\times} {\rm d}\Tc/{\rm d}P_{\parallel [100]}+{\rm d}\Tc/{\rm d}P_{\parallel [001]}$, 
indicates that at least either ${\rm d}\Tc/{\rm d}P_{\parallel [100]}$ or ${\rm d}\Tc/{\rm d}P_{\parallel [001]}$ should be negative. 
In addition, using the Ehrenfest relation, the uniaxial-pressure coefficients of $\Tc$ can be estimated
from the discontinuity at $\Tc$ in the longitudinal elastic modulus obtained from ultrasonic measurements.\cite{Okuda2002JPSJ} 
The evaluated uniaxial-pressure coefficients are $\frac{1}{\Tc}\frac{{\rm d}\Tc}{{\rm d}P_{\parallel [100]}}=-(0.85\pm 0.05)$~GPa$^{-1}$ and
$\frac{1}{\Tc}\frac{{\rm d}\Tc}{{\rm d}P_{\parallel [001]}}=+(0.7\pm0.2)$~GPa$^{-1}$.
Note that although the coefficients have nearly the same magnitude, they have opposite sign.

On theoretical grounds, this estimation is supported at least in a qualitative fashion. 
The Fermi surface of $\214$ consists of three nearly-cylindrical sheets called the $\alpha$, $\beta$, and $\gamma$ bands
derived mainly from Ru-4$d$ electrons hybridized with O-2$p$ electrons.
The active band $\gamma$ originates mainly from the $d_{xy}$ orbital, 
in contrast to the passive bands $\alpha$ and $\beta$, originating mainly from the $d_{zx}$ and $d_{yz}$ orbitals.
Under uniaxial pressure along the $c$ axis, 
the energy levels of the $d_{zx}$ and $d_{yz}$ orbitals increase due to the crystal field effect, 
while the $d_{xy}$ orbital's does not.\cite{Nomura2002JPSJ-2}
Consequently, electrons are transfered from the $\alpha$ and $\beta$ bands to the $\gamma$ band. 
This causes the Fermi level to approach a van Hove singularity in the $\gamma$ band, 
increasing its density of states of the Fermi surface. \cite{Nomura2002JPSJ-2}
Therefore, $\Tc$ is expected to increase for out-of-plane pressure.
The effect for in-plane pressure is opposite; $\Tc$ would be expected to decrease.

In the context of the above discussion, the present results are striking. 
Although the uniaxial-pressure effect along the $c$ axis obtained in this study is consistent with the above predictions,
the sign of the in-plane uniaxial-pressure effect is opposite to that expected.
Moreover, the magnitude of the in-plane pressure effect is substantially greater than that of the out-of-plane pressure effect. 
As a prominent feature presented in Fig.~\ref{com}, 
while the shielding fraction continues to increase up to the maximum pressure reached,
$T_\mathrm{c1}$ and $T_\mathrm{c2}$ for in-plane pressures saturate at approximately 3.3~K, 
very close to the onset $\Tc$ of the 3-K phase at 0~GPa ($\sim 3.5$~K).\cite{Kittaka2009JPSJ}
This suggests that the onset temperature itself does not change significantly under uniaxial pressure.

In this eutectic system, a large strain is expected to develop due to 
the differences in the thermal contraction and lattice compressibility between $\214$ and Ru
as the crystal cools to room temperature or below after solidification from the melt.\cite{Ando1999JPSJ}
A plausible origin of the enhancement of $\Tc$ is the in-plane rotation of the RuO$_6$ octahedra 
to release the strain at interfaces between $\214$ and Ru, 
as suggested by Sigrist and Monien in ref. 10. 
This RuO$_6$ rotation could be easily induced by in-plane pressure 
which affects the Ru-Ru lattice constant in the $ab$ plane directly. 
In fact, the $\Sigma_3$ soft mode at the (0.5 0.5 0) zone boundary, 
corresponding to the RuO$_6$ rotation about the $c$ axis, was observed 
in inelastic neutron scattering experiments on $\214$.\cite{Braden1998PRB}
A lattice distortion of this kind would reduce the dispersion of the $\gamma$ band and 
cause an increase in the $\Tc$ in pure $\214$ regions via an enhancement of the density of states at the Fermi level.\cite{Nomura2002JPSJ-2}
Uniaxial pressure in the presence of the interface may couple strongly to this instability,
possibly leading to lattice distortions over an extended spatial region.

The temperature dependence of the upper critical field $\Hc2$ of the 3-K phase 
exhibits upturn behavior below $\sim$~2.3~K and above $\sim$~0.2~T \cite{Yaguchi2003PRB}, 
which allows the length scale $\delta$ of the $\214$ region with enhanced $\Tc$ 
to be estimated \cite{Matsumoto2003JPSJ}.
With decreasing temperature, the nucleation region of the superconductivity
shrinks as the coherence length $\xi_{ab}(T)=\sqrt{\hbar/(2e\mu_0H_{\mathrm{c2} \parallel c}(T))}$ changes.
When $\xi_{ab}(T) \lesssim \delta$ is satisfied,  the nucleation of the superconductivity 
is well confined in the region with a finite width of $\delta$ where $\Tc$ is enhanced.
Confinement in this region is characterized by shorter local coherence length corresponding to enhanced $\Tc$, and 
leads to enhanced $H_{\mathrm{c2} \parallel c}(T)$ compared to that at higher temperatures.
In the present model, the onset field of the upturn will be 
around the temperature at which $\xi_{ab}(T) \sim \delta$ is satisfied. 

Matsumoto $et$ $al$. \cite{Matsumoto2003JPSJ} used Ginzburg Landau formalism 
to analyze properties of the 3-K phase in magnetic fields, 
similar to Sigrist and Monien's theory \cite {Sigrist2001JPSJ}, 
and assumed a region surrounding a Ru inclusion with enhanced $\Tc$ 
that extends away into $\214$ part with a finite width of $\delta$.
A fit to the experimental data\cite{Yaguchi2003PRB} using their theory 
yields $\delta \approx 200$ $\rm{\AA}$. \cite{Matsumoto2003JPSJ}

In the present study, 
the apparent shielding fraction was revealed to be as high as 40\% at an in-plane pressure of 0.4~GPa. 
Although it may be overestimated due to a demagnetization effect,
the actual shielding fraction is estimated to be at least 10\% using the calculated demagnetization factor.
Therefore, the marked enhancement of the shielding fraction indicates that $\delta$ develops to nearly 1 $\muup$m
because the distance between adjacent Ru inclusions is on the average about 10 $\muup$m.
Surprisingly, the present results suggest that $\delta$ at 0.4~GPa becomes larger by a factor of about a hundred than $\delta$ at 0~GPa.

In summary, we have investigated the effect of uniaxial pressure on superconductivity in the Sr$_2$RuO$_4$-Ru eutectic system.
Uniaxial pressures in all of the applied directions strongly enhance the superconducting volume fraction of the 3-K phase,
but hardly enhance its onset temperature. 
Contrary to the expectations deduced from the Ehrenfest relation for pure $\214$, 
the effect of in-plane pressure is greater than that of out-of-plane pressure.
Surprisingly, at 0.4~GPa for $P_{\parallel [110]}$, the shielding fraction at 1.8~K exceeds 40\%.
This remarkable enhancement of the shielding fraction indicates that 
3-K phase superconductivity penetrates deep into the bulk of the $\214$ region by uniaxial stress.
This striking magnitude of the effect as well as its anisotropy may help resolving the origin of 3-K phase superconductivity itself.
We propose that uniaxial pressure stabilizes a lattice distortion near the interfaces between $\214$ and Ru,
leading to a strong enhancement of the superconducting volume fraction of the 3-K phase.
These findings urge the uniaxial pressure effect on pure $\214$ to be investigated as well,
although such investigation is technically more difficult because pure $\214$ crystals are cleaved easily. 

We thank K. Takizawa, N. Takeshita, M. Sigrist, Y. Machida, S. Yonezawa, K. Ishida, and D. C. Peets for their support and valuable discussions. 
This work is supported by a Grant-in-Aid for the Global COE program ``The Next Generation of Physics, Spun from Universality and Emergence'' 
from the Ministry of Education, Culture, Sports, Science, and Technology (MEXT) of Japan.
It is also supported by Grants-in-Aid for Scientific Research from MEXT and from the Japan Society for the Promotion of Science (JSPS).
S. K. is supported as a JSPS Research Fellow.

\end{document}